\documentclass[twocolumn]{jpsj3}

\title{%
Unified Formula for Stationary Josephson Current
in Planar Graphene Junctions
}

\author{%
Yositake Takane
}

\inst{%
Department of Quantum Matter,
Graduate School of Advanced Science and Engineering,\\
Hiroshima University, Higashihiroshima, Hiroshima 739-8530, Japan
}

\recdate{ \hspace{50mm} }

\abst{%
The stationary Josephson current in a ballistic graphene system
is theoretically studied with focus on a planar junction
consisting of a monolayer graphene sheet on top of which
a pair of superconducting electrodes is deposited.
To characterize such a planar junction, we employ two parameters:
the coupling strength between the graphene sheet and
the superconducting electrodes,
and a potential drop induced in the graphene sheet
by direct contact with the electrodes.
We derive a general formula for the Josephson current by taking
these parameters into account in addition to other basic parameters,
such as temperature and chemical potential.
The resulting formula applies to a wide range of parameters
and reproduces previously reported results in certain limits.
}

\kword{%
Josephson current, graphene junction, ballistic regime
}

\begin{document}
\sloppy
\maketitle

\section{Introduction}

For more than a decade, the Josephson effect~\cite{josephson}
in a superconductor-graphene-superconductor (SGS) junction has attracted
considerable theoretical~\cite{wakabayashi,titov,moghaddam,gonzalez,
black-schaffer,hayashi,hagymasi,alidoust,takane1,takane2,rakyta,yang,
pellegrino}
and experimental~\cite{lee1,heersche,sato,du,ojeda,kanda,tomori,jeong,
komatsu,choi,mizuno,calado,shalom,borzenets,nanda,park,lee2,jang} interest.
In most studies, researchers attempted to observe
how the stationary Josephson current is affected by the unique band structure
of a graphene sheet,~\cite{novoselov,castro_neto}
in which the conduction and valence bands touch conically
at $K_{+}$ and $K_{-}$ points in the Brillouin zone (the Dirac points).
In early experiments, such an attempt was not easy to succeed
because the graphene sheet used to fabricate an SGS junction is not
sufficiently clean, thus, electron motion cannot be ballistic in it.
However, the encapsulation technique of a graphene sheet enables us
to fabricate a nearly ideal SGS junction~\cite{lee1,shalom,borzenets,calado,
nanda,park,lee2,jang} in which the electron motion is ballistic.
In such an SGS junction, the unique band structure of a graphene sheet
should manifest itself in various features of the Josephson current.
To elucidate such features, a general theoretical description of
the Josephson current is highly desirable.

Here, we briefly review a theoretical study by
Titov and Beenakker,~\cite{titov}
which serves as a starting point of the theoretical approach
to the Josephson effect in an SGS junction.
The SGS junction considered in Ref.~\citen{titov} is depicted in Fig. 1,
where two superconductors $\rm S_{1}$ ($L/2 \le x$)
and $\rm S_{2}$ ($x \le -L/2$) of width $W$ are placed with
separation $L$ on top of a clean monolayer graphene sheet
with the condition of $L \ll W$.
In Ref.~\citen{titov}, it is assumed that electron states
in the graphene sheet are described by a massless Dirac equation,
and that the carrier doping in the covered region of $L/2 \le |x|$
is described by an effective potential
of a negative constant $-U$.~\cite{comment1}
In Ref.~\citen{titov}, it is also assumed that
the superconducting proximity effect on the graphene sheet is described by
an energy-independent effective pair potential $\Delta_{\rm eff}$,
which is constant in the covered region ($L/2 \le |x|$)
and vanishes in the uncovered region  ($|x| \le L/2$).
The important parameters characterizing the Josephson current
in this model are $L$, $U$, and $\Delta_{\rm eff}$,
in addition to temperature $T$ and chemical potential $\mu$.
By taking the limit of $U \to \infty$,
the authors of Ref.~\citen{titov} derived a formula for
the Josephson current  at $T = 0$ in the short junction limit of $L \ll \xi$,
where $\xi$ is the superconducting coherence length.
The formula, given in Eq.~(19) of Ref.~\citen{titov},
applies to $0 \le \mu$~\cite{comment2}
under the condition of $T = 0$, $L \ll \xi$, and $U \to \infty$.

The assumption of $\Delta_{\rm eff}$ being energy-independent
was examined in Refs.~\citen{takane1} and \citen{takane2} to improve
the description of the superconducting proximity effect.
In Refs.~\citen{takane1} and \citen{takane2},
the proximity effect is described by treating the coupling between
the graphene sheet and the superconducting electrodes
in terms of a tunneling Hamiltonian.~\cite{mcmillan,takane3}
Instead of using $\Delta_{\rm eff}$, this approach adopts a parameter $\Gamma$
that controls the strength of the tunnel coupling, enabling us
to take into account the energy dependence of the effective pair potential.
It is shown that the resulting formula, given in Eq.~(53) of
Ref.~\citen{takane2}, cohesively describes various behaviors of
the Josephson critical current $I_{c}$ as a function of $T$
observed in a set of samples.~\cite{park}
In particular, it succeeds in describing the unusual $T$ dependence of $I_{c}$
in an SGS junction with a relatively weak coupling.
A drawback of this formula is that its application is restricted to
the case of $\mu$ being sufficiently away from the Dirac point.
This is ascribed to a quasiclassical approximation used in its derivation.

The purpose of this study is to give a general formula for
the stationary Josephson current through a monolayer graphene sheet,
which can be applied to a wide range of parameters.
To do so, we adopt the model used in Ref.~\citen{takane2}
and derive a general formula for the Josephson current
without relying on a quasiclassical approximation.
The resulting formula applies to arbitrary
$T$, $\mu$, $L$, $U$, and $\Gamma$,~\cite{comment3}
and reproduces the formulas of Refs.~\citen{titov} and \citen{takane2}
in certain limits.
The paper is organized as follows.
In Sect.~2, we describe the model for the SGS junction
and introduce a thermal Green's function.
In Sect.~3, we construct the thermal Green's function
and then derive a general formula for the Josephson current.
In Sect.~4, we show that the resulting formula reproduces
the results of Refs.~\citen{titov} and \citen{takane2}
in certain limits.
In Sect.~5, the behavior of the Josephson critical current is numerically
studied in a short junction limit.
Section 6 is devoted to a summary.
We set $k_{\rm B} = \hbar = 1$ throughout the paper.

\section{Model and Thermal Green's Function}

We consider an SGS junction of monolayer graphene as depicted in Fig.~1.
We adopt a model described in Ref.~\citen{takane2}
and then introduce a thermal Green's function that is convenient
for the subsequent analysis of the Josephson current,
\begin{figure}[btp]
\begin{center}
\includegraphics[height=4.0cm]{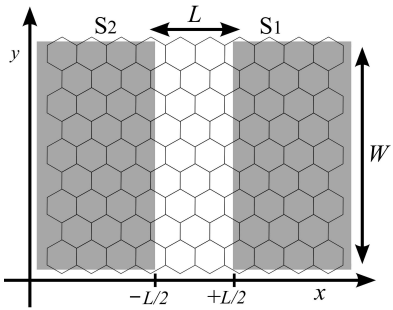}
\end{center}
\caption{Josephson junction consisting of a monolayer graphene sheet
on which two superconductors ${\rm S}_{1}$ and ${\rm S}_{2}$
of width $W$ are deposited with separation $L$.
}
\end{figure}

In Fig.~1, two superconductors ${\rm S}_{1}$ and ${\rm S}_{2}$ of width $W$
are placed with separation $L$ on top of a clean monolayer graphene sheet,
where ${\rm S}_{1}$ and ${\rm S}_{2}$ respectively occupy the regions of
$L/2 \le x$ and of $x \le -L/2$.
We assume that the pair potential is given by
\begin{align}
  \Delta (x)
  = \left\{ \begin{array}{cc}
               \Delta e^{i \varphi/2}
               & (L/2 < x) \\
               0 & (|x| < L/2) \\
               \Delta e^{-i \varphi/2}
               & (x < -L/2) ,
            \end{array}
    \right.
\end{align}
where $\varphi$ serves as the phase difference
between the two superconducting electrodes.

Let us assume that
the coupling of the graphene sheet and the superconductors
is described by a tunneling Hamiltonian.
The resulting proximity effect on the graphene sheet is described
by a self-energy~\cite{mcmillan,takane3} [see Eq.~(\ref{eq:self-ene})].
The coupling with the superconductors also induces carrier doping
in the graphene sheet; the carrier density in the covered region of
$L/2 < |x|$ becomes higher than that in the uncovered region of $|x| < L/2$.
We describe this by adding the effective potential of a negative constant $-U$
only in the covered region,~\cite{titov}
resulting in the renormalization of the chemical potential $\mu$:
\begin{align}
  \tilde{\mu}
  = \left\{ \begin{array}{cc}
               \mu & (|x| < L/2) \\
               \mu + U  & (L/2 < |x|) .
            \end{array}
    \right.
\end{align}

Let us turn to the electron states in the graphene sheet.
Low-energy states appear in the two valleys located at
the $K_{+}$ and $K_{-}$ points in the Brillouin zone,
where the wave vector corresponding to
the $K_{\pm}$ point is given by $\mib{K}_{\pm} = \pm(2\pi/a)(2/3,0)$
with $a$ being the lattice constant of the graphene sheet.
Within the effective mass approximation,
the low-energy states in the $K_{\pm}$ valley are described by
the effective Hamiltonian $H_{\pm}$
defined by~\cite{wallace,mcclure,slonczewski}
\begin{align}
          \label{eq:H-pm}
  H_{\pm} & = \left( \begin{array}{cc}
                       -\tilde{\mu} & \gamma k_{\mp} \\
                       \gamma k_{\pm} & -\tilde{\mu}
                     \end{array}
              \right) ,
\end{align}
where $k_{\pm} = k_{x} \pm i k_{y}$ with
$k_{x} = - i \partial_{x}$ and $k_{y} = - i \partial_{y}$.
The $2 \times 2$ form of $H_{\pm}$ reflects the fact that
the unit cell of a hexagonal lattice contains A and B sites,
and $\gamma$ is given by $\gamma = (\sqrt{3}/2)\gamma_{0}a$,
where $\gamma_{0}$ represents the nearest-neighbor transfer
integral.~\cite{wallace,mcclure,slonczewski}

In the presence of the superconducting proximity effect,
we need to treat electron and hole states
taking their coupling into account.
The simplest way to do this is to employ
a Bogoliubov--de Gennes equation:
\begin{align}
   H_{\rm BdG}\left( \begin{array}{c}
                       \Psi_{e} \\ \Psi_{h}
                     \end{array}
              \right)
   = \epsilon \left( \begin{array}{c}
                       \Psi_{e} \\ \Psi_{h}
                     \end{array}
              \right) ,
\end{align}
where $\Psi_{e}$ and $\Psi_{h}$ are respectively
the electron and hole wavefunctions, and
the $4 \times 4$ Hamiltonian for the $K_{+}$ valley
is given by~\cite{beenakker}
\begin{align}
 H_{\rm BdG}
   = \left( \begin{array}{cc}
              H_{+}
              & \Delta_{\rm eff}(x) \sigma_{0} \\
              \Delta_{\rm eff}(x)^{*} \sigma_{0}
              & -H_{+}
            \end{array}
     \right)
\end{align}
with $\sigma_{0} = {\rm diag}(1,1)$.
Here, $\Delta_{\rm eff}(x)$ is the effective pair potential, which is usually
assumed to be an energy-independent constant in the covered region.
This widely accepted assumption for $\Delta_{\rm eff}$ is justified only when
the coupling between the graphene sheet and the superconducting electrodes
is sufficiently strong.~\cite{takane2,takane3}
To cope with arbitrary coupling strength,
we employ the tunneling Hamiltonian model proposed by McMillan~\cite{mcmillan}
instead of assuming the energy-independent pair potential.
The approach of McMillan is reformulated in Ref.~\citen{takane3}
in the form specific to a hybrid graphene system.

We introduce the $4 \times 4$ thermal Green's function
$G(\mib{r},\mib{r}';\omega)$ with $\omega = (2n+1)\pi T$, which obeys
\begin{align}
       \label{eq:Green}
  \left( i \omega \tau^{0} - H - \Sigma
  \right) G(\mib{r},\mib{r}';\omega)
  = \tau^{0} \delta(\mib{r}-\mib{r}') ,
\end{align}
where $H = {\rm diag}(H_{+}, -H_{+})$
and $\tau^{0} = {\rm diag}(1,1,1,1)$.
The self-energy $\Sigma$, representing the proximity effect
mediated by quasiparticle tunneling, is given by~\cite{takane1,takane3}
\begin{align}
    \label{eq:self-ene}
  \Sigma
  &  = \frac{-\Gamma \theta\left(|x|-\frac{L}{2}\right)}
            {\sqrt{\Delta^{2}+\omega^{2}}}
       \left( \begin{array}{cc}
                i \omega & \Delta(x) \\
                \Delta(x)^{*} & i \omega
              \end{array}
       \right)
       \otimes \sigma_{0} ,
\end{align}
where $\Gamma$ represents the strength of the tunnel coupling and
$\theta(x)$ is the Heaviside step function.
The off-diagonal elements are regarded as an energy-dependent
effective pair potential,
while the diagonal elements describe the renormalization of a
quasiparticle energy.
Here and hereafter, we restrict our consideration to quasiparticle states
in the $K_{+}$ valley because those in the $K_{-}$ valley
equivalently contribute to the Josephson current.
A brief comment on $G(\mib{r},\mib{r}';\omega)$ is given in Appendix A.

\section{Formulation}

We derive a general formula for the Josephson current by using
an analytical expression of the thermal Green's function on the basis of
the argument originally given by Ishii~\cite{ishii1,ishii2} and later developed
by Furusaki and Tsukada.~\cite{furusaki1,furusaki2,furusaki3}

Hereafter, we restrict our attention to the regime of electron doping:
$0 \le \mu < \mu+U$.
Assuming that our system is translationally invariant in the $y$ direction,
we perform the Fourier transformation:
\begin{align}
  G(x,x';q,\omega)
 = \int d(y-y') e^{-iq(y-y')}
   G(\mib{r},\mib{r}';\omega) ,
\end{align}
which we explicitly express as
\begin{align}
  G(x,x';q,\omega)
  = \left( \begin{array}{cc}
             g(x,x';q,\omega) & f'(x,x';q,\omega) \\
             f^{\dagger}(x,x';q,\omega) & g'(x,x';q,\omega)
           \end{array}
    \right) .
\end{align}
Note that we need to treat only
$g(x,x';q,\omega)$ and $f^{\dagger}(x,x';q,\omega)$.
Let us consider them in the uncovered region of $|x| < L/2$.
It is convenient to define the wave numbers in the $x$-direction as
\begin{align}
  k_{e} & = {\rm sgn}_{\omega}
            \sqrt{\left(\frac{\mu+i\omega}{\gamma}\right)^{2}-q^{2}} ,
     \\
  k_{h} & = {\rm sgn}_{\omega}
             \sqrt{\left(\frac{\mu-i\omega}{\gamma}\right)^{2}-q^{2}} ,
\end{align}
where ${\rm Im}\{k_{e}\} > 0$ and ${\rm Im}\{k_{h}\} < 0$,
and ${\rm sgn}_{\omega}$ represents the sign of $\omega$.
It is also convenient to introduce
\begin{align}
 & e^{\pm i\phi_{e}} = \frac{\gamma(k_{e}\pm i q)}{\mu + i\omega} ,
     \\
 & e^{\pm i\phi_{h}} = \frac{\gamma(k_{h}\pm i q)}{\mu - i\omega} .
\end{align}
This is equivalent to defining
\begin{align}
 & \cos\phi_{e} = \frac{\gamma k_{e}}{\mu + i\omega} , \hspace{5mm}
   \sin\phi_{e} = \frac{\gamma q}{\mu + i\omega} ,
     \\
 & \cos\phi_{h} = \frac{\gamma k_{h}}{\mu - i\omega} , \hspace{5mm}
   \sin\phi_{h} = \frac{\gamma q}{\mu - i\omega} .
\end{align}

If $\mu$ is sufficiently away from the Dirac point, the Josephson current is
carried by propagating modes.
References~\citen{takane1} and \citen{takane2} focus on this case,
in which $k_{e}$ and $k_{h}$, respectively, can be approximated as
Eqs.~(\ref{eq:ke-HCL}) and (\ref{eq:kh-HCL}), reproducing the result of
a quasiclassical Green's function approach.~\cite{takane1,takane2}
Contrastingly, if $\mu$ is very near the Dirac point, the Josephson current
is carried by evanescent modes.
In this study, we treat these two different cases as well as
an intermediate case in a unified manner.

A general solution of $g(x,x';q,\omega)$ is written as
\begin{align}
     \label{eq:g-exp}
  g(x,x';q,\omega)
 & =  \left[ - \frac{i}{v_{e}}\theta(x-x')+ c_{++} \right]
      e^{ik_{e}(x-x')}\Lambda^{++}_{e}
          \nonumber \\
 &  \hspace{0mm}
    + \left[ - \frac{i}{v_{e}}\theta(x'-x)+ c_{--} \right]
      e^{-ik_{e}(x-x')}\Lambda^{--}_{e}
          \nonumber \\
 &  \hspace{-15mm}
    + c_{+-} e^{ik_{e}(x+x')}\Lambda^{+-}_{e}
    + c_{-+} e^{-ik_{e}(x+x')}\Lambda^{-+}_{e} ,
\end{align}
where $v_{e} = \gamma \cos\phi_{e}$ and
\begin{align}
  \Lambda^{++}_{e}
 &   = \frac{1}{2}  
       \left( \begin{array}{cc}
                       1 & e^{-i \phi_{e}} \\
                       e^{i \phi_{e}} & 1
              \end{array}
       \right) ,
        \\
  \Lambda^{--}_{e}
 &   = \frac{1}{2}  
       \left( \begin{array}{cc}
                       1 & -e^{i \phi_{e}} \\
                       -e^{-i \phi_{e}} & 1
              \end{array}
       \right) ,
        \\
  \Lambda^{+-}_{e}
 &   = \frac{1}{2}  
       \left( \begin{array}{cc}
                       e^{-i \phi_{e}} & -1 \\
                       1 & -e^{i \phi_{e}}
              \end{array}
       \right) ,
        \\
  \Lambda^{-+}_{e}
 &   = \frac{1}{2}  
       \left( \begin{array}{cc}
                       e^{i \phi_{e}} & 1 \\
                       -1 & -e^{-i \phi_{e}}
              \end{array}
       \right) .
\end{align}
A general solution of $f^{\dagger}(x,x';q,\omega)$ is written as
\begin{align}
     \label{eq:f-exp}
 & f^{\dagger}(x,x';q,\omega)
        \nonumber \\
 & \hspace{-7mm}
    =  d_{++}e^{i(k_{h}x-k_{e}x')}\Lambda^{++}_{h}
    + d_{--}e^{-i(k_{h}x-ik_{e}x')}\Lambda^{--}_{h}
          \nonumber \\
 &  \hspace{-7mm}
    + d_{+-} e^{i(k_{h}x+k_{e}x')}\Lambda^{+-}_{h}
    + d_{-+} e^{-i(k_{h}x+k_{e}x')}\Lambda^{-+}_{h} ,
\end{align}
where
\begin{align}
  \Lambda^{++}_{h}
 &   = \frac{1}{2}  
       \left( \begin{array}{cc}
                       e^{-\frac{i}{2}(\phi_{h}-\phi_{e})}
                         & e^{-\frac{i}{2}(\phi_{h}+\phi_{e})} \\
                       e^{\frac{i}{2}(\phi_{h}+\phi_{e})}
                         & e^{\frac{i}{2}(\phi_{h}-\phi_{e})}
              \end{array}
       \right) ,
        \\
  \Lambda^{--}_{h}
 &   = \frac{1}{2}  
       \left( \begin{array}{cc}
                       e^{\frac{i}{2}(\phi_{h}-\phi_{e})}
                         & -e^{\frac{i}{2}(\phi_{h}+\phi_{e})} \\
                       -e^{-\frac{i}{2}(\phi_{h}+\phi_{e})}
                         & e^{-\frac{i}{2}(\phi_{h}-\phi_{e})}
              \end{array}
       \right) ,
        \\
  \Lambda^{+-}_{h}
 &   = \frac{1}{2}  
       \left( \begin{array}{cc}
                       e^{-\frac{i}{2}(\phi_{h}+\phi_{e})}
                         & -e^{-\frac{i}{2}(\phi_{h}-\phi_{e})} \\
                       e^{\frac{i}{2}(\phi_{h}-\phi_{e})}
                         & -e^{\frac{i}{2}(\phi_{h}+\phi_{e})}
              \end{array}
       \right) ,
        \\
  \Lambda^{-+}_{h}
 &   = \frac{1}{2}  
       \left( \begin{array}{cc}
                       e^{\frac{i}{2}(\phi_{h}+\phi_{e})}
                         & e^{\frac{i}{2}(\phi_{h}-\phi_{e})} \\
                       -e^{-\frac{i}{2}(\phi_{h}-\phi_{e})}
                         & -e^{-\frac{i}{2}(\phi_{h}+\phi_{e})}
              \end{array}
       \right) .
\end{align}

The Josephson current is formally expressed as
\begin{align}
        \label{eq:I_start_0}
  I(\varphi)
  = 4W \int_{-\infty}^{+\infty}\frac{dq}{2\pi}T\sum_{\omega}
    {\rm tr}\left\{ j_{x}g(x;q,\omega)
            \right\} ,
\end{align}
where the factor $4$ comes from the spin and valley degeneracies,
the current operator $j_{x}$ is defined by
\begin{align}
   j_{x}
 & = e\gamma
     \left( \begin{array}{cc}
              0 & 1 \\
              1 & 0
            \end{array}
     \right),
\end{align}
and $g(x;q,\omega) \equiv \frac{1}{2}[g(x,x-0;q,\omega)+g(x,x+0;q,\omega)]$.
Substituting Eq.~(\ref{eq:g-exp}) into Eq.~(\ref{eq:I_start_0}),
we obtain
\begin{align}
        \label{eq:I_start}
  I(\varphi)
  = 4e W \int_{-\infty}^{+\infty}
    dq v_{e}
    T\sum_{\omega} \left(c_{++}(\varphi)-c_{--}(\varphi)\right) .
\end{align}

The unknown coefficients $c_{++}$ and $c_{--}$ are determined by
a boundary condition at $x = \pm L/2$ for $g(x,x';q,\omega)$ and
$f^{\dagger}(x,x';q,\omega)$, which we briefly describe below.
By solving the Bogoliubov--de Gennes equation in the covered region
of $L/2 \le |x|$ (see Appendix B),
we find a relationship between the electron wavefunction $\Psi_{e}$
and the hole wavefunction $\Psi_{h}$, which is expressed by using
\begin{align}
    \label{eq:def-tilde-ome}
  \tilde{\omega}
  & = \left(1+\frac{\Gamma}{\sqrt{\omega^{2}+\Delta^{2}}} \right)\omega ,
          \\
    \label{eq:def-tilde-Del}
  \tilde{\Delta}
  & = \frac{\Gamma}{\sqrt{\omega^{2}+\Delta^{2}}} \Delta ,
          \\
  \Omega
  & = {\rm sgn}_{\omega} \sqrt{\tilde{\omega}^{2}+\tilde{\Delta}^{2}} ,
\end{align}
and $\chi$ defined by
\begin{align}
       \label{eq:def-chi}
  e^{\pm i\chi} = \frac{\gamma (p \pm i q)}{\mu + U}
\end{align}
with
\begin{align}
       \label{eq:def-p}
  p = {\rm sgn}_{\omega}\sqrt{\left(\frac{\mu + U}{\gamma}\right)^{2}-q^{2}} ,
\end{align}
where $U$ is assumed to be the largest energy scale in our model.
Let $\Psi^{+}_{e}$ ($\Psi^{+}_{h}$) and $\Psi^{-}_{e}$ ($\Psi^{-}_{h}$)
be respectively the right-going and left-going components
of $\Psi_{e}$ ($\Psi_{h}$).
At $x = \pm L/2$, they satisfy
\begin{align}
        \label{eq:BC-condition}
     \Psi^{\pm}_{e} = B(\pm L/2) \Psi^{\pm}_{h}
\end{align}
with
\begin{align}
  B(\pm L/2) & = -i \frac{e^{\pm i\varphi/2}}{\tilde{\Delta}\cos\chi}
     \nonumber \\
             & \hspace{-10mm} \times
             \left(
               \begin{array}{cc}
                 \tilde{\omega}\cos\chi \mp i\Omega\sin\chi & \pm \Omega \\
                 \pm \Omega & \tilde{\omega}\cos\chi \pm i\Omega \sin\chi
               \end{array}
             \right) .
\end{align}
The derivation of Eq.~(\ref{eq:BC-condition}) is outlined in Appendix B.
Equation~(\ref{eq:BC-condition}), serving as the boundary condition,
gives a set of coupled equations:
\begin{align}
 & \left( -\frac{i}{v_{e}}+c_{++}\right)e^{ik_{e}\frac{L}{2}}\Lambda^{++}_{e}
     +c_{-+}e^{-ik_{e}\frac{L}{2}}\Lambda^{-+}_{e}
          \nonumber \\
 & = B(L/2)
     \left(  d_{++}e^{ik_{h}\frac{L}{2}}\Lambda^{++}_{h}
            +d_{-+}e^{-ik_{h}\frac{L}{2}}\Lambda^{-+}_{h}
     \right) ,
           \\
 &   c_{--}e^{-ik_{e}\frac{L}{2}}\Lambda^{--}_{e}
     +c_{+-}e^{ik_{e}\frac{L}{2}}\Lambda^{+-}_{e}
          \nonumber \\
 & = B(L/2)
     \left(  d_{--}e^{-ik_{h}\frac{L}{2}}\Lambda^{--}_{h}
            +d_{+-}e^{ik_{h}\frac{L}{2}}\Lambda^{+-}_{h}
     \right) ,
           \\
 & \left( -\frac{i}{v_{e}}+c_{--}\right)e^{ik_{e}\frac{L}{2}}\Lambda^{--}_{e}
     +c_{+-}e^{-ik_{e}\frac{L}{2}}\Lambda^{+-}_{e}
          \nonumber \\
 & = B(-L/2)
     \left(  d_{--}e^{ik_{h}\frac{L}{2}}\Lambda^{--}_{h}
            +d_{+-}e^{-ik_{h}\frac{L}{2}}\Lambda^{+-}_{h}
     \right) ,
           \\
 &   c_{++}e^{-ik_{e}\frac{L}{2}}\Lambda^{++}_{e}
     +c_{-+}e^{ik_{e}\frac{L}{2}}\Lambda^{-+}_{e}
          \nonumber \\
 & = B(-L/2)
     \left(  d_{++}e^{-ik_{h}\frac{L}{2}}\Lambda^{++}_{h}
            +d_{-+}e^{ik_{h}\frac{L}{2}}\Lambda^{-+}_{h}
     \right) .
\end{align}

Solving these equations, we obtain
\begin{align}
     \label{eq:c_++--}
 c_{++}(\varphi) = c_{--}(-\varphi)
 = -ie^{-i\frac{\varphi}{2}} \frac{\zeta}{2 v_{e}\Xi}
\end{align}
with
\begin{align}
 \zeta
 & =  e^{i(k_{e}-k_{h})\frac{L}{2}}
            \nonumber \\
 & \hspace{0mm}
      \times
       \Biggl[ \tilde{\omega}\cos\chi
              \cos\left(\frac{\phi_{e}+\phi_{h}}{2}\right)
            \nonumber \\
 & \hspace{7mm}
            - \Omega
              \left(  \cos\left(\frac{\phi_{e}-\phi_{h}}{2}\right)
                    - \sin\chi \sin\left(\frac{\phi_{e}+\phi_{h}}{2}\right)
              \right)
      \Biggr]
            \nonumber \\
 & \hspace{0mm}
      \times 
      \Biggl[  -i\tilde{\omega}
               \cos\chi
               \cos\left(\frac{\phi_{e}+\phi_{h}}{2}\right)
               \sin\left((k_{e}-k_{h})\frac{L}{2}+\frac{\varphi}{2}\right)
            \nonumber \\
 & \hspace{7mm}
             + \Omega
               \left(  \cos\left(\frac{\phi_{e}-\phi_{h}}{2}\right)
                     - \sin\chi \sin\left(\frac{\phi_{e}+\phi_{h}}{2}\right)
               \right)
            \nonumber \\
 & \hspace{10mm}
               \times
               \cos\left((k_{e}-k_{h})\frac{L}{2}+\frac{\varphi}{2}\right)
      \Biggr]
            \nonumber \\
 & - e^{i(k_{e}+k_{h})\frac{L}{2}}
            \nonumber \\
 & \hspace{0mm}
      \times
       \Biggl[ \tilde{\omega}\cos\chi
              \cos\left(\frac{\phi_{e}-\phi_{h}}{2}\right)
            \nonumber \\
 & \hspace{7mm}
            - \Omega
              \left(  \sin\chi \cos\left(\frac{\phi_{e}-\phi_{h}}{2}\right)
                    - \sin\left(\frac{\phi_{e}+\phi_{h}}{2}\right)
              \right)
      \Biggr]
            \nonumber \\
 & \hspace{0mm}
      \times 
      \Biggl[-i\tilde{\omega}
               \cos\chi
               \sin\left(\frac{\phi_{e}-\phi_{h}}{2}\right)
               \sin\left((k_{e}+k_{h})\frac{L}{2}+\frac{\varphi}{2}\right)
            \nonumber \\
 & \hspace{7mm}
             + \Omega
               \left(  \sin\chi \cos\left(\frac{\phi_{e}-\phi_{h}}{2}\right)
                     - \sin\left(\frac{\phi_{e}+\phi_{h}}{2}\right)
               \right)
            \nonumber \\
 & \hspace{10mm}
               \times
               \cos\left((k_{e}+k_{h})\frac{L}{2}+\frac{\varphi}{2}\right)
      \Biggr] ,
         \\
 \Xi & =  \frac{1}{2}
          \Biggl[  \tilde{\omega}^{2}\cos^{2}\chi
                   \cos^{2}\left(\frac{\phi_{e}+\phi_{h}}{2}\right)
            \nonumber \\
 & \hspace{5mm}
                 + \Omega^{2}
                   \left(  \cos\left(\frac{\phi_{e}-\phi_{h}}{2}\right)
                         - \sin\chi\sin\left(\frac{\phi_{e}+\phi_{h}}{2}\right)
                   \right)^{2}
          \Biggr]
            \nonumber \\
 & \hspace{10mm}
          \times
          \cos\left((k_{e}-k_{h})L\right)
            \nonumber \\
 & \hspace{1mm}
        - i\tilde{\omega}\Omega
          \cos\chi \cos\left(\frac{\phi_{e}+\phi_{h}}{2}\right)
            \nonumber \\
 & \hspace{10mm}
          \times
          \left(  \cos\left(\frac{\phi_{e}-\phi_{h}}{2}\right)
                - \sin\chi \sin\left(\frac{\phi_{e}+\phi_{h}}{2}\right)
          \right)
            \nonumber \\
 & \hspace{10mm}
          \times
          \sin\left((k_{e}-k_{h})L\right)
            \nonumber \\
 & \hspace{1mm}
       -  \frac{1}{2}
          \Biggl[ \tilde{\omega}^{2}\cos^{2}\chi
                  \sin^{2}\left(\frac{\phi_{e}-\phi_{h}}{2}\right)
            \nonumber \\
 & \hspace{5mm}
                + \Omega^{2}
                  \left(  \sin\chi\cos\left(\frac{\phi_{e}-\phi_{h}}{2}\right)
                        - \sin\left(\frac{\phi_{e}+\phi_{h}}{2}\right)
                  \right)^{2}
          \Biggr]
            \nonumber \\
 & \hspace{10mm}
          \times
          \cos\left((k_{e}+k_{h})L\right)
            \nonumber \\
 & \hspace{1mm}
        + i\tilde{\omega}\Omega \cos\chi
          \sin\left(\frac{\phi_{e}-\phi_{h}}{2}\right)
            \nonumber \\
 & \hspace{10mm}
          \times
          \left(  \sin\chi \cos\left(\frac{\phi_{e}-\phi_{h}}{2}\right)
                - \sin\left(\frac{\phi_{e}+\phi_{h}}{2}\right)
          \right) 
            \nonumber \\
 & \hspace{10mm}
          \times
          \sin\left((k_{e}+k_{h})L\right)
            \nonumber \\
 & \hspace{1mm}
        + \frac{1}{2}
          \tilde{\Delta}^{2}\cos^{2}\chi\cos\phi_{e}\cos\phi_{h} \cos\varphi .
\end{align}

Substituting Eq.~(\ref{eq:c_++--}) into Eq.~(\ref{eq:I_start}),
we finally obtain
\begin{align}
        \label{eq:I_final}
    I(\varphi)
    = \frac{eW}{\pi} \int_{-\infty}^{+\infty}
      dq T\sum_{\omega}
      \frac{\tilde{\Delta}^{2}\cos^{2}\chi
            \cos\phi_{e}\cos\phi_{h}}{\Xi} \sin\varphi .
\end{align}
This is the central result of this paper.
Using this general formula, we can numerically calculate
the Josephson current in an SGS junction for arbitrary parameters.

\section{Limiting Cases}

We show that Eq.~(\ref{eq:I_final}) reproduces the previous results
of Refs.~\citen{titov} and \citen{takane2} in certain limits.
In this sense, we can regard it as a unified formula
for the stationary Josephson current in a planar SGS junction.

\subsection{Short junction limit}

Let us focus on the short junction limit of $L \ll \xi$,
where $\xi \equiv \gamma/(2\pi \Delta_{0})$
is the superconducting coherence length
with $\Delta_{0}$ being the pair potential at $T = 0$.
In this limit, $\omega$ in $k_{e}$ and $k_{h}$ can be ignored.~\cite{titov}
This results in $k_{e} = k_{h} = k$ for $\gamma q < \mu$
and $k_{e} = -k_{h} = k$ for $\mu < \gamma q$, where
\begin{align}
    k = {\rm sgn}_{\omega} \sqrt{\left(\frac{\mu}{\gamma}\right)^{2}-q^{2}}
\end{align}
with ${\rm Im}\{k\} \ge 0$.
Accordingly, we find $\phi_{e} = \phi_{h} = \phi$ for $\gamma q < \mu$
and $\phi_{e} = {\rm sgn}_{q}\,\pi - \phi_{h} = \phi$ for $\mu < \gamma q$,
where
\begin{align}
 e^{\pm i \phi} = \frac{\gamma \left(k \pm i q\right)}{\mu}
\end{align}
and ${\rm sgn}_{q}$ represents the sign of $q$.
Hence, $\Xi$ in Eq.~(\ref{eq:I_final}) is reduced to $\Xi_{\rm SJL}$
for $\gamma q < \mu$ and $-\Xi_{\rm SJL}$ for $\mu < \gamma q$, where
\begin{align}
  \Xi_{\rm SJL}
 & = \tilde{\omega}^{2}
     \left( \cos^{2}\chi\cos^{2}\phi +(\sin\chi-\sin\phi)^{2}\sin^{2}kL \right)
             \nonumber \\
 & \hspace{+2mm}
   + \tilde{\Delta}^{2}
     \Bigl( \cos^{2}\chi\cos^{2}\phi +(\sin\chi-\sin\phi)^{2}\sin^{2}kL
             \nonumber \\
 & \hspace{16mm}
            -\cos^{2}\chi\cos^{2}\phi\sin^{2}\frac{\varphi}{2} \Bigr) .
\end{align}
We obtain the expression of the Josephson current
in the short junction limit:
\begin{align}
        \label{eq:I_short}
    I_{\rm SJL}(\varphi)
    = \frac{eW}{\pi} \int_{-\infty}^{+\infty}dq 
      T\sum_{\omega}
      \frac{\tau(q)\tilde{\Delta}^{2}\sin\varphi}
           {\tilde{\omega}^{2}
            + \tilde{\Delta}^{2}
              \bigl[1-\tau(q)\sin^{2}\frac{\varphi}{2}\bigr]} ,
\end{align}
where
\begin{align}
  \tau(q) = \frac{\cos^{2}\chi \cos^{2}\phi}
                 {\cos^{2}\chi \cos^{2}\phi
                  +(\sin\chi-\sin\phi)^{2}\sin^{2}kL} .
\end{align}

Let us restrict our consideration to the strong coupling limit of
$\Gamma \to \infty$,
where $\tilde{\omega}/\tilde{\Delta}$ can be replaced with $\omega/\Delta$.
After performing the summation over $\omega$, we find
\begin{align}
        \label{eq:I_short-strong}
    I_{\rm SJL}(\varphi)
  &  = \frac{e\Delta W}{2\pi}\int_{-\infty}^{\infty}dq
       \frac{\tau(q)\sin\varphi}{\sqrt{1-\tau(q)\sin^{2}\frac{\varphi}{2}}}
              \nonumber \\
  & \hspace{5mm} \times
       \tanh\left(\frac{\Delta}{2T}\sqrt{1-\tau(q)\sin^{2}\frac{\varphi}{2}}
           \right) .
\end{align}
At $T = 0$, this expression is reduced to Eq.~(19) of Ref.~\citen{titov}
in the limit of $U \to \infty$,
where $\cos^{2}\chi =1$ and $\sin\chi = 0$.~\cite{comment4}
Equation~(\ref{eq:I_short-strong}) should be regarded as
an extension of the result of Kulik and Omel'yanchuk.~\cite{kulik}

\subsection{High-carrier-density limit}

Let us next consider the high-carrier-density limit of
$\gamma/L$, $\Delta_{0} \ll \mu$.
In this limit, we can approximate that
\begin{align}
    \label{eq:ke-HCL}
  k_{e} & = k + \frac{\mu}{\gamma^{2}k}i\omega ,
      \\
    \label{eq:kh-HCL}
  k_{h} & = k - \frac{\mu}{\gamma^{2}k}i\omega ,
\end{align}
and $\phi_{e} = \phi_{h} = \phi$.
Hence, $\Xi$ in Eq.~(\ref{eq:I_final}) is reduced to
\begin{align}
  \Xi_{\rm HCL}
 & = \frac{1}{2}\left[  \tilde{\omega}^{2}\cos^{2}\chi\cos^{2}\phi
                      + \Omega^{2}\left(1-\sin\chi\sin\phi\right)^{2}
                \right]
             \nonumber \\
 & \hspace{10mm} 
                \times
                \cosh\left(\frac{2\omega L}{v_{x}}\right)
             \nonumber \\
 & \hspace{2mm}
   + \tilde{\omega}\Omega\cos\chi\cos\phi\left(1-\sin\chi\sin\phi\right)
     \sinh\left(\frac{2\omega L}{v_{x}}\right)
            \nonumber \\
 & \hspace{2mm}
   - \frac{1}{2}\Omega^{2}\left(\sin\chi-\sin\phi\right)^{2}\cos2kL
            \nonumber \\
 & \hspace{2mm}
   + \frac{1}{2}\tilde{\Delta}^{2}\cos^{2}\chi\cos^{2}\phi\cos\varphi ,
\end{align}
where $v_{x} = \gamma \cos\phi$.
We obtain the expression of the Josephson current
in the high-carrier-density limit:
\begin{align}
        \label{eq:I_HDR}
    I_{\rm HCL}(\varphi)
    = \frac{eW}{\pi}\int_{-\infty}^{\infty}dq
      T\sum_{\omega}
      \frac{\tilde{\Delta}^{2}\cos^{2}\chi\cos^{2}\phi}
           {\Xi_{\rm HCL}}
      \sin\varphi .
\end{align}
This expression is equivalent to Eq.~(53) of Ref.~\citen{takane2},
derived by using a quasiclassical Green's function approach.

\section{Numerical Result}

We focus on the short junction limit of $L \ll \xi$ with heavy doping
in the covered region (i.e., $\gamma/L$, $\Delta_{0} \ll U$),
which is particularly important in actual experiments.
The Josephson critical current $I_{c}$ defined by
\begin{align}
  I_{c} = \underset{\varphi}{\rm max} \{I(\varphi)\}
\end{align}
is numerically calculated as a function of $T$
in the high-carrier-density case of $\mu/\Delta_{0} = 200$
and the low-carrier-density case of $\mu/\Delta_{0} = 1$.
The critical current is also calculated as a function of $\mu$.
In every case, we set $\Gamma/\Delta_{0} = 1$, $20$, and $2000$
with $U/\Delta_{0} = 4000$.
The following parameters are employed:
$L = 200 \ {\rm nm}$, $W = 4 \ \mu{\rm m}$, $\gamma_{0} = 2.8 \ {\rm eV}$,
$a = 0.246 \ {\rm nm}$, and $\Delta_{0} = 120 \ \mu{\rm eV}$.
The coherence length is estimated as $\xi = 2.4 \, \mu{\rm m}$,
which is much larger than $L$.
The behavior of $I_{c}$ in the short junction limit
is fully described by Eq.~(\ref{eq:I_short}).
The amplitude of the pair potential is determined by the gap equation
\begin{align}
  1 = \lambda_{\rm int} 
  \int_{0}^{\epsilon_{\rm D}} {\rm d}\epsilon
  \tanh\left(\frac{\sqrt{\epsilon^{2}+\Delta^{2}}}{2T}\right)/
  \sqrt{\epsilon^{2}+\Delta^{2}} ,
\end{align}
where $\lambda_{\rm int}$ is the dimensionless interaction constant,
and the Debye energy is chosen as $\epsilon_{\rm D}/\Delta_{0} = 200$.

Figure~2 shows $I_{c}$ in the high-carrier-density case
of $\mu /\Delta_{0} = 200$ normalized by
\begin{align}
  I_{0} = e\Delta_{0}\frac{\mu W}{\pi \gamma}
\end{align}
as a function of $T/T_{\rm c}$ with $\Gamma/\Delta_{0} = 1$, $20$, and $2000$.
The $I_{\rm c}$ curve is convex upward for $\Gamma/\Delta_{0} = 20$ and $2000$,
whereas it becomes convex downward for $\Gamma/\Delta_{0} = 1$.
Figure~3 shows $I_{c}$ in the low-carrier-density case
of $\mu /\Delta_{0} = 1$ normalized by
\begin{align}
  I_{0} = e\Delta_{0}\frac{W}{\pi L}
\end{align}
as a function of $T/T_{\rm c}$ with $\Gamma/\Delta_{0} = 1$, $20$, and $2000$.
The $I_{c}$ curve also shows a crossover
from convex upward to convex downward with decreasing $\Gamma/\Delta_{0}$.

\begin{figure}[bpt]
\begin{center}
\includegraphics[height=5.0cm]{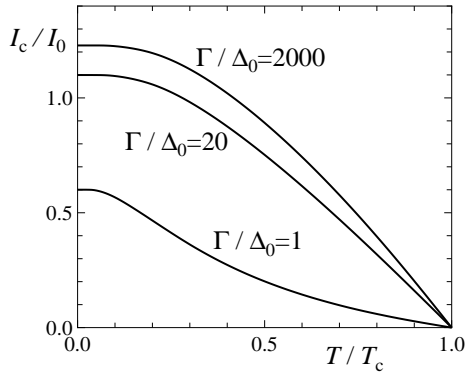}
\end{center}
\caption{Critical current $I_{c}$
normalized by $I_{0} = e\Delta_{0}\frac{\mu W}{\pi \gamma}$
as a function of $T/T_{\rm c}$
at $\mu/\Delta_{0} = 200$ for $\Gamma/\Delta_{0} = 1$, $20$, and $2000$.
}
\end{figure}
\begin{figure}[bpt]
\begin{center}
\includegraphics[height=5.0cm]{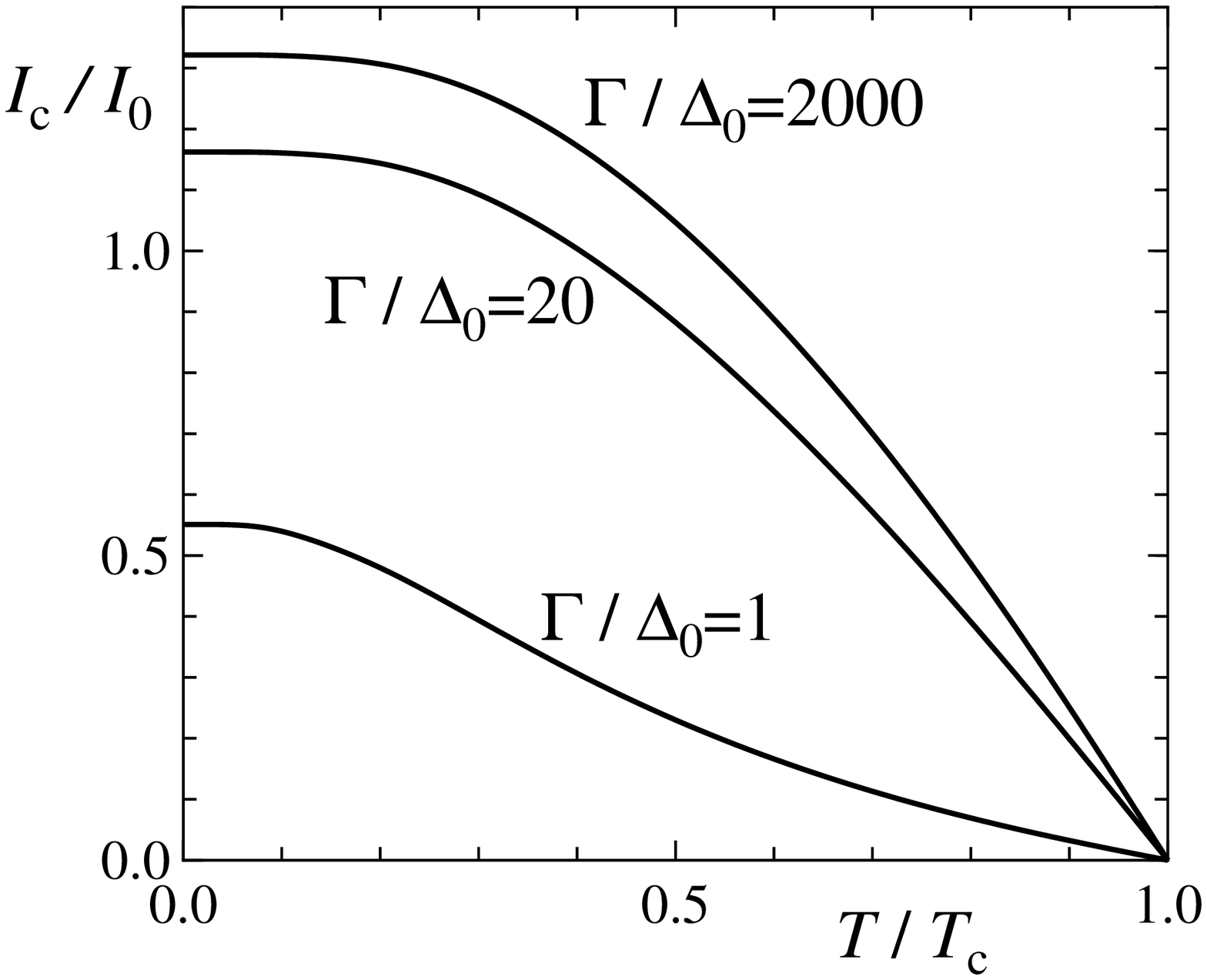}
\end{center}
\caption{Critical current $I_{c}$ normalized by
$I_{0} = e\Delta_{0}\frac{W}{\pi L}$ as a function of $T/T_{\rm c}$
at $\mu/\Delta_{0} = 1$ for $\Gamma/\Delta_{0} = 1$, $20$, and $2000$.
}
\end{figure}
\begin{figure}[btp]
\begin{center}
\includegraphics[height=5.0cm]{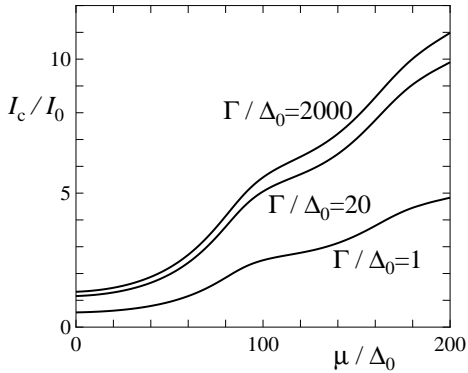}
\end{center}
\caption{Critical current $I_{c}$
normalized by $I_{0} = e\Delta_{0}\frac{W}{\pi L}$ as a function of $\mu$
at $T/T_{\rm c} = 0.01$ for $\Gamma/\Delta_{0} = 1$, $20$, and $2000$.
}
\end{figure}

As noted in the previous section, Eq. (\ref{eq:I_short}) reproduces
Eq.~(19) of Ref.~\citen{titov} at $T = 0$ if $\Gamma$ and $U$ are
sufficiently large.
Thus, the resulting $I_{c}$ in the strong coupling case of
$\Gamma/\Delta_{0} = 2000$ is expected to reproduce the corresponding results
of Ref.~\citen{titov}.
Indeed, for $\Gamma/\Delta_{0} = 2000$, $I_{c}/I_{0}$ in the case of
$\mu/\Delta_{0} = 200$ is $1.228$ at $T=0$,
which is quantitatively consistent with Eq.~(22) of Ref.~\citen{titov}.
Similarly, $I_{c}/I_{0}$ in the case of $\mu/\Delta_{0} = 1$
is $1.321$ at $T=0$,
which is also quantitatively consistent with Eq.~(21) of Ref.~\citen{titov}.

Figure~4 shows $I_{c}$ as a function of $\mu$
for $\Gamma/\Delta_{0} = 1$, $20$, and $2000$ at $T/T_{\rm c} = 0.01$,
where $I_{c}$ is normalized by $I_{0} = e\Delta_{0}\frac{W}{\pi L}$.

\section{Summary}

Adopting a simple model of SGS junctions, we derive a general formula
for the stationary Josephson current.
The resulting formula contains $T$, $\mu$, $L$, $U$, and $\Gamma$
as important parameters
and is applicable to an arbitrary set of these parameters,~\cite{comment3}
where $T$ is temperature, $\mu$ is chemical potential,
$L$ is the separation between two superconducting electrodes,
$U$ controls the carrier doping in the graphene sheet,
and $\Gamma$ represents the coupling strength between the graphene
sheet and the superconducting electrodes.
We show that it reproduces the formula of Ref.~\citen{titov} in the limit of
$L \ll \xi$, $U \to \infty$, and $\Gamma \to \infty$ at $T = 0$.
We also show that it is reduced to the formula of Ref.~\citen{takane2}
in the limit of $\gamma/L$, $\Delta_{0} \ll \mu$,
where $\gamma$ is the velocity of an electron in a graphene sheet
and $\Delta_{0}$ is the pair potential at $T = 0$.

\section*{Acknowledgment}

This work was supported by JSPS KAKENHI Grant Number JP18K03460.

\appendix

\section{Components of Green's function}

The thermal Green's function $G(\mib{r},\mib{r}';\omega)$ is described by
the effective Hamiltonian $\tilde{H}$ defined by
\begin{align}
      \label{eq:BdG-H}
   \tilde{H} =
   \left( \begin{array}{cccc}
            -\tilde{\mu}(x) & \gamma k_{-} & -\tilde{\Delta}(x) & 0 \\
            \gamma k_{+} & -\tilde{\mu}(x) & 0 & -\tilde{\Delta}(x) \\
            -\tilde{\Delta}^{*}(x) & 0 & \tilde{\mu}(x) & -\gamma k_{-} \\
            0 & -\tilde{\Delta}^{*}(x) & -\gamma k_{+} & \tilde{\mu}(x)
          \end{array}
   \right) ,
\end{align}
which possesses the particle-hole symmetry:~\cite{beenakker}
\begin{align}
   \Theta^{-1} \tilde{H} \Theta = - \tilde{H} ,
\end{align}
where
\begin{align}
  \Theta = \left( \begin{array}{cc}
                    0 & -\vartheta \\
                    \vartheta & 0
                  \end{array}
           \right)
\end{align}
with $\vartheta = -i\sigma_{y}K$.
Here, $\sigma_{y}$ is the $y$ component of Pauli matrix and $K$ denotes
a complex conjugate operator.

Let us express the thermal Green's function as
\begin{align}
  G(\mib{r},\mib{r}';\omega)
  = \left( \begin{array}{cc}
             g(\mib{r},\mib{r}';\omega) & f'(\mib{r},\mib{r}';\omega) \\
             f^{\dagger}(\mib{r},\mib{r}';\omega) & g'(\mib{r},\mib{r}';\omega)
           \end{array}
    \right) .
\end{align}
Using a spectral representation with the help of the particle-hole symmetry,
we can represent $g'(\mib{r},\mib{r}';\omega)$ and
$f'(\mib{r},\mib{r}';\omega)$ in terms of $g(\mib{r},\mib{r}';\omega)$ and
$f^{\dagger}(\mib{r},\mib{r}';\omega)$, respectively.
Here, we present only the final results,
\begin{align}
   g'(\mib{r},\mib{r}';\omega)
   & = - \vartheta^{-1} g(\mib{r},\mib{r}';\omega) \vartheta ,
        \\
   f'(\mib{r},\mib{r}';\omega)
   & = \vartheta^{-1} f^{\dagger}(\mib{r},\mib{r}';\omega) \vartheta .
\end{align}

\section{Derivation of Boundary Condition}

By solving the Bogoliubov--de Gennes equation in a Matsubara
representation, we present wavefunctions in the covered region
of $L/2 \le |x|$.
The boundary condition, given in Eq.~(\ref{eq:BC-condition}), is
straightforwardly obtained from the resulting wavefunctions.
The Bogoliubov--de Gennes equation in the region of $L/2 \le x$ 
is written as
\begin{align}
      \label{eq:BdG-covered}
 &  \left( i\tilde{\omega}\tau^{0} - \tilde{H} \right)
    \left( \begin{array}{c}
             \Psi_{e} \\ \Psi_{h}
           \end{array}
    \right) = 0 ,
\end{align}
where $\tilde{H}$ is given in Eq.~(\ref{eq:BdG-H}), and $\tilde{\mu}$ and
$\tilde{\Delta}(x)$ in it should read as $\tilde{\mu} = \mu + U$
and $\tilde{\Delta}(x)=\tilde{\Delta}e^{i\frac{\varphi}{2}}$, respectively.
Hereafter, we assume that $U$ is much larger than $\Delta_{0}$.

It is convenient to define $\kappa$ as
\begin{align}
 \kappa
  = \frac{\mu+U}{\gamma^{2}p}\Omega .
\end{align}
By using the wave number $q$ in the transverse direction in addition to
$\kappa$, $p$, and $\chi$ (the latter two are defined in the text),
the right-going wave function $\Psi^{+} =^{t}\!(\Psi_{e}^{+},\Psi_{h}^{+})$
and the left-going wavefunction $\Psi^{-}=^{t}\!(\Psi_{e}^{-},\Psi_{h}^{-})$
in the region of $L/2 \le x$ are respectively expressed as
\begin{align}
     \left(
     \begin{array}{c}
       \Psi_{e}^{+} \\ \Psi_{h}^{+}
     \end{array}
     \right)
 & = e^{ipx-\kappa x +iqy}
     \left(
     \begin{array}{c}
       e^{-i\frac{\chi}{2}}
       \frac{\tilde{\omega}+\Omega}{\tilde{\Delta}} \\
       e^{i\frac{\chi}{2}}
       \frac{\tilde{\omega}+\Omega}{\tilde{\Delta}} \\
       i e^{-i\frac{\chi}{2}}e^{-i\frac{\varphi}{2}} \\
       i e^{i\frac{\chi}{2}}e^{-i\frac{\varphi}{2}}
    \end{array}
    \right) ,
          \\
     \left(
     \begin{array}{c}
       \Psi_{e}^{-} \\ \Psi_{h}^{-}
     \end{array}
     \right)
 & = e^{-ipx-\kappa x +iqy}
     \left(
     \begin{array}{c}
       e^{i\frac{\chi}{2}}
       \frac{\tilde{\omega}-\Omega}{\tilde{\Delta}} \\
       -e^{-i\frac{\chi}{2}}
       \frac{\tilde{\omega}-\Omega}{\tilde{\Delta}} \\
       i e^{i\frac{\chi}{2}}e^{-i\frac{\varphi}{2}} \\
       -i e^{-i\frac{\chi}{2}}e^{-i\frac{\varphi}{2}}
    \end{array}
    \right) .
\end{align}
From these equations, we can easily derive the boundary condition
[i.e., Eq.~(\ref{eq:BC-condition})] at $x = L/2$.

The Bogoliubov--de Gennes equation in the region of $x \le -L/2$
is equivalent to Eq.~(\ref{eq:BdG-covered}) if we set
$\tilde{\Delta}(x)=\tilde{\Delta}e^{-i\frac{\varphi}{2}}$.
The right-going wavefunction $\Psi^{+} =^{t}\!(\Psi_{e}^{+},\Psi_{h}^{+})$
and the left-going wavefunction $\Psi^{-}=^{t}\!(\Psi_{e}^{-},\Psi_{h}^{-})$
in the region of $x \le -L/2$ are respectively expressed as
\begin{align}
     \left(
     \begin{array}{c}
       \Psi_{e}^{+} \\ \Psi_{h}^{+}
     \end{array}
     \right)
 & = e^{ipx + \kappa x + iqy}
     \left(
     \begin{array}{c}
       e^{-i\frac{\chi}{2}}
       \frac{\tilde{\omega}-\Omega}{\tilde{\Delta}} \\
       e^{i\frac{\chi}{2}}
       \frac{\tilde{\omega}-\Omega}{\tilde{\Delta}} \\
       i e^{-i\frac{\chi}{2}}e^{i\frac{\varphi}{2}} \\
       i e^{i\frac{\chi}{2}}e^{i\frac{\varphi}{2}}
    \end{array}
    \right) ,
          \\
     \left(
     \begin{array}{c}
       \Psi_{e}^{-} \\ \Psi_{h}^{-}
     \end{array}
     \right)
 & = e^{-ipx + \kappa x + iqy}
     \left(
     \begin{array}{c}
       e^{i\frac{\chi}{2}}
       \frac{\tilde{\omega}+\Omega}{\tilde{\Delta}} \\
       -e^{-i\frac{\chi}{2}}
       \frac{\tilde{\omega}+\Omega}{\tilde{\Delta}} \\
       i e^{i\frac{\chi}{2}}e^{i\frac{\varphi}{2}} \\
       -i e^{-i\frac{\chi}{2}}e^{i\frac{\varphi}{2}}
    \end{array}
    \right) .
\end{align}
From these equations, we can easily derive the boundary condition
[i.e., Eq.~(\ref{eq:BC-condition})] at $x = -L/2$.


\begin{thebibliography}{99}
\bibitem{comment1} The assumption of $U$ being a piecewise constant
is widely accepted in the literature
but not easy to justify in an explicit manner.
There is indirect evidence supporting it.
Indeed, Eq.~(53) of Ref.~\citen{takane2} derived under this assumption
properly describes a set of experimental results reported
in Ref.~\citen{park}.
\bibitem{comment3} The restriction of $0<\mu(\mu+U)$ is implicitly assumed.
See also comments given above~\cite{comment2}.
\bibitem{comment4} The integration over $q$ can be replaced with
the summation over $q_{n} = \pi\left(n+\frac{1}{2}\right)/W$ as
$\frac{W}{2\pi}\int_{-\infty}^{\infty}dq \to \sum_{n=0}^{\infty}$.
\end{thebibliography}
\end{document}